\documentclass[11pt]{article}

\usepackage{graphicx}                  
\usepackage{dcolumn}                   
\usepackage{bm}

\usepackage{jheppub}                   

\makeatletter
\gdef\@fpheader{\ }                    
\makeatother

\usepackage{graphicx}
\usepackage{dcolumn}
\usepackage{bm}
\usepackage{float}

\usepackage{amsmath,amssymb}           
\usepackage{amscd}                     
\usepackage{epsfig}                    
\usepackage[matrix,arrow]{xy}          
\usepackage{xspace}                    
\usepackage{stmaryrd}                  
\usepackage{slashed}
\usepackage{tikz}
\usepackage{hyperref}

\DeclareSymbolFont{bbold}{U}{bbold}{m}{n}
\DeclareSymbolFontAlphabet{\mathbbold}{bbold}

\newcommand{\bq}{\begin{equation}}
\newcommand{\eq}{\end{equation}}
\newcommand{\bea}{\begin{eqnarray}}
\newcommand{\eea}{\end{eqnarray}}

\newcommand{\dd}{\mathrm{d}}
\newcommand{\ee}{\mathrm{e}}
\newcommand{\ii}{\mathrm{i}}

\newcommand{\der}{\partial}

\newcommand{\bbZ}{\mathbb{Z}}
\newcommand{\bbR}{\mathbb{R}}

\DeclareMathOperator{\SU}{\mathit{SU}}

\DeclareMathOperator{\USp}{\mathit{USp}}
\DeclareMathOperator{\Symp}{\mathit{USp}}
\DeclareMathOperator{\Spin}{\mathit{Spin}}

\DeclareMathOperator{\Cliff}{Cliff}

\newcommand{\rep}[1]{\mathbf{#1}}
\newcommand{\repp}[2]{(\rep{#1}, \rep{#2})}
\newcommand{\id}{\mathbbold{1}}

\newcommand{\Gs}[1]{\Gamma(#1)}

\newcommand{\Dgen}{{D}}

\newcommand{\LC}{\nabla}

\newcommand{\GenRic}{R^{\scriptscriptstyle 0}}

\newcommand{\GenS}{R}

\newcommand{\proj}[1]{\times_{#1}}

\DeclareMathOperator{\Edd}{\mathit{E_{d(d)}}}

\DeclareMathOperator{\dHd}{\mathit{\tilde{H}_d}}

\newcommand{\tF}{{\tilde{F}}}

\newcommand{\Tgen}{W}		 	
\newcommand{\Tint}{\Tgen_{\text{int}}}
\newcommand{\TTint}{T_{\text{int}}}

\newcommand{\CC}{\text{c.c.}}

\notoc

\title{Generalised Structures for $\mathcal{N}=1$ AdS Backgrounds}

\author[a]{Andr\'e Coimbra,}
\emailAdd{andre.coimbra@itp.uni-hannover.de}
\author[b]{Charles Strickland-Constable}
\emailAdd{charles.strickland-constable@cea.fr}

\affiliation[a]{Institut f\"ur Theoretische Physik \& 
  Center for Quantum Engineering and Spacetime Research,\\
  Leibniz Universit\"at Hannover, Appelstra{\ss}e 2, 
  30167 Hannover, Germany }

\affiliation[b]{Institut de physique th\'eorique, 
	Universit\'e Paris Saclay, CEA, CNRS, F-91191 Gif-sur-Yvette, France}

\subheader{\textrm{IPhT-T15/35} \\ \textrm{ITP-UH-04/15}}

\abstract{We expand upon a claim made in a recent paper [arXiv:1411.5721] that generic minimally supersymmetric AdS backgrounds of warped flux compactifications of Type II and M theory can be understood as satisfying a straightforward weak integrability condition in the language of $E_{d(d)}\times\mathbb{R}^+$ generalised geometry. Namely, they are spaces admitting a generalised $G$-structure set by the Killing spinor and with constant singlet generalised intrinsic torsion.
\vfill}

\begin{document}  
 
\maketitle


\section{Introduction}
\label{sec:intro}

In the context of the study of flux compactifications of string theory, the problem of describing possible supersymmetric Anti-de Sitter solutions has acquired central importance with the discovery of the AdS/CFT correspondence~\cite{Maldacena:1997re}. Substantial progress has been achieved in understanding the geometry of such backgrounds of M theory and Type II using the tools of $G$-structures (see for instance~\cite{Gauntlett:2002sc,Gauntlett:2002fz, Martelli:2003ki,Behrndt:2004bh, Lukas:2004ip,Gauntlett:2004zh, Lust:2004ig, Gauntlett:2005ww, Gabella:2012rc}) and generalised geometry (for example in~\cite{Minasian:2006hv,Gabella:2009wu,Gabella:2010cy, Apruzzi:2013yva, Apruzzi:2014qva,Apruzzi:2015zna}).

Combining both approaches, it was shown in a recent work~\cite{CSW4} that it is possible to characterise fully generic minimally supersymmetric compactifications to $D\geq4$-dimensional Minkowski space by a novel integrability condition, formulated in the language of $\Edd\times\bbR^+$ generalised geometry~\cite{chris,PW,CSW2} (throughout $d=11-D$ and $\Edd$ is the real split form of the rank $d$ exceptional Lie group). Concretely, it was proven that the Killing spinor equations constrain precisely the intrinsic torsion of the generalised $G$-structure defined by the Killing spinor on the generalised tangent bundle. In other words, the compactification space must be the generalised analogue of a special holonomy manifold, as setting the supersymmetry variations of the fermions to zero is equivalent to demanding that the intrinsic torsion vanishes. 

Let us briefly recall the key features of exceptional generalised geometry for the description of supersymmetric backgrounds of Type II and M theory. Given a Riemannian spin manifold $M$ with $d\leq 7$ dimensions for M theory or $d-1\leq 6$ dimensions for Type II,
\begin{itemize}
\item[\textbf{--}] we introduce the generalised tangent bundle $E$, which enlarges the usual tangent bundle to also accommodate the symmetries of the supergravity gauge fields~\cite{PW,chris};
\item[\textbf{--}] the bundle $E$ has the structure group $\Edd\times\bbR^+$, so we can construct generalised tensors associated to $\Edd\times\bbR^+$ representations~\cite{PW,CSW2};
\item[\textbf{--}] there is a differential structure on $E$ described by the exceptional Dorfman bracket, which generates both diffeomorphisms and gauge transformations~\cite{PW};
\item[\textbf{--}] we can define generalised connections $\Dgen_V$ to take derivatives of generalised tensors along generalised vectors $V\in\Gs{E}$, and a natural notion of generalised torsion tensor determined by the Dorfman bracket~\cite{CSW2};
\item[\textbf{--}] there exists a generalised metric on $E$, whose components unify all the bosonic fields in the theory, and which is invariant under transformations of the maximal compact subgroup $H_d\subset\Edd$ which generalises orthogonal transformations~\cite{PW,chris};
\item[\textbf{--}] the double cover of this local group, $\dHd$, can be realised as a subgroup of the Clifford algebra $\Cliff(d;\bbR)$, and so we can think of spinors as representations of $\dHd$~\cite{CSW2,CSW3};
\item[\textbf{--}] the existence of a globally non-vanishing spinor $\epsilon$ on $M$ defines a subgroup $G\subset \dHd$ which stabilises it, ie. a generalised $G$-structure~\cite{PW,GO,GO2,CSW3};
\item[\textbf{--}] there exists a torsion-free generalised connection compatible with $G$ if and only if $M$ is the internal space of a minimally supersymmetric Minkowski flux background~\cite{CSW4}.
\end{itemize}

Some subtleties aside, this last result was obtained simply by looking at the relevant groups and representations, which are summarised in table~\ref{table}.

\begin{table}[H]
\centering
\begin{tabular}{llll}
   $d$ & $\dHd$ & $G$ & $\Tint\simeq$ KSEs \\
   \hline 
   7 & $\SU(8)$  
      & $\SU(7)$ & $\rep{1} + \rep{7} + \rep{21}+ \rep{35} + \CC$\\
   6 & $ \Symp(8)$ & $\Symp(6)$ 
      & $2\cdot\rep{1} + 2\cdot\rep{6} + 2\cdot\rep{14} + \rep{14'} + \CC $\\
   5 & $\Symp(4)^2$ & $\Symp(2)\cdot\Symp(4)$
      & $2\cdot\repp{1}{4} + 2\cdot\repp{2}{4} + \CC $ \\
   4 & $\Symp(4)$ & $\Symp(2)$ 
      & $4\cdot\rep{1} + 5\cdot\rep{2} + 2\cdot\rep{3} + \CC$
\end{tabular}
\caption{Generalised spin group $\dHd$; stabiliser group $G\subset\dHd$ of the Killing spinor; and the space of intrinsic torsions $\Tint$ of the generalised $G$-structure, which was proven to match the decomposition of the Killing spinor equations (KSEs). 
}
\label{table}
\end{table}

It was further claimed in~\cite{CSW4} that generic minimally supersymmetric AdS compactifications could be similarly described, now by keeping certain singlet components of the generalised intrinsic torsion as a non-zero constant, corresponding to the cosmological constant $\Lambda$. We thus have that
\begin{quote}
\textit{The minimally supersymmetric $D\geq 4$ AdS backgrounds are in one-to-one correspondence with weak generalised $G$ holonomy spaces, with singlet torsion given by the cosmological constant and where  $G=\SU(7), \USp(6), \USp(2)\times\USp(4), \USp(2)$ in dimensions $D=4,5,6,7$ respectively.}
\end{quote} 
In the following we will clarify this statement and demonstrate it explicitly.


\section{AdS backgrounds}

We consider generic supersymmetric flux compactifications of M theory and Type II string theory to four- and higher-dimensional  AdS space. This means we have the warped metric ansatz
\begin{equation}
\label{eq:g}
   \dd s^2 = \ee^{2A}\dd s^2(\text{AdS}_D) + \dd s^2(M) , 
\end{equation}
with $D\geq 4$ and where the warp factor $A$ is a scalar function of the internal coordinates. The internal space $M$ is a spin manifold with Riemannian metric $g$, of dimension $d$ in M theory and $d-1$ in Type II. To match the conventions of~\cite{CSW2,CSW3}, we take $A=\Delta$ in M theory, so that $A=\Delta+\frac{1}{3}\phi$ in Type II, where $\phi$ is the dilaton, and the metric is in the string frame. For the fluxes we keep only the components consistent with the $D$-dimensional AdS symmetry. Fermion fields are set to zero and we work in the supergravity limit $\alpha'=0$.

\subsection{Killing spinor equations}

In a supersymmetric AdS space one has Killing spinors $\eta$ which must satisfy
\begin{equation}\label{eq:ads-kse}
\begin{aligned}
	\LC_\mu \eta_{\pm} &= \tfrac12 \ee^{\pm 2\ii \theta} \Lambda \gamma_\mu  \eta_{\mp}, &\quad \text{in } D = 4,\\
	\LC_\mu \eta^A &= \tfrac12 M^A{}_B \Lambda \gamma_\mu  \eta^B, &\quad \text{in } D = 5,\\
	\LC_\mu \eta_\pm^A &= \tfrac12 (N^{\pm1}){}^A{}_B \Lambda \gamma_\mu  \eta_\mp^A, &\quad \text{in } D = 6,\\
	\LC_\mu \eta^A &= \tfrac12 \Lambda \gamma_\mu  \eta^A, &\quad \text{in } D = 7,\\
\end{aligned}
\end{equation}
where $\LC$ is the Levi--Civita connection in $\text{AdS}_D$, $A,B$ are the $\SU(2)$ indices of the symplectic Majorana and Majorana-Weyl spinors in $D=5,7$ and $D=6$ respectively, and $\pm$ subscripts denote chirality in even dimensions under the top gamma matrix $\gamma^{(D)}$.\footnote{For more details on conventions on Clifford algebras, intertwiners, spinor decompostions, etc. please refer to appendix \textbf{B} of~\cite{CSW3}. Here, for convenience, we choose a representation in which $(\gamma_m)^T = (\gamma_m)^* = - \gamma_m$ for $d=6,7$.} 

In order to write these as R-symmetry covariant equations we have included a constant arbitrary phase $\theta$ in $D=4$, a constant $2\times2$ traceless Hermitian matrix, which squares to the identity, $M^A{}_B$ rotating the sympletic Majorana spinors in $D=5$, and for $D=6$, the matrix $N^A{}_B$ which is a constant element of $\SU(2)$. Usually these equations are written with particular values for $\theta, M,N$. One could, for example, rotate the spinors $\eta$ by R-symmetry transformations to chose $\theta = 0$, $M = \sigma^3$ and $N = \id$. Doing so explicitly breaks the Minkowski R-symmetry group and allows us to directly obtain the surviving R-symmetry in AdS. For instance, of the full $U(1)$ R-symmetry in $D=4$ we see that only a residual $\mathbb{Z}_2$ would remain, while in $D=7$ the equation is actually invariant under the full $\SU(2)$, so the R-symmetry stays the same as in flat space.

We must now tensor these with the internal Killing spinors to obtain a supersymmetric solution of the full higher-dimensional theory.  For concreteness we will focus on the M theory description, though our analysis covers the Type II cases straightforwardly~\cite{CSW2,CSW3,CSW4}.

Given spinors $\epsilon$ on the internal space $M$, we construct an eleven-dimensional spinor $\varepsilon^-$ as 
\begin{equation}
\begin{aligned}
	\varepsilon^- &= \eta_+ \otimes \epsilon + \eta_- \otimes \epsilon^*, &\quad \text{in } D = 4,\\
	\varepsilon^- &= \epsilon_{AB} \, \eta^A \otimes \epsilon^B, &\quad \text{in } D = 5,\\
	\varepsilon^- &= \epsilon_{AB} \, \eta_+^A \otimes \epsilon_1^B 
		+ \epsilon_{AB} \, \eta_-^A \otimes \epsilon_2^B, &\quad \text{in } D = 6,\\
	\varepsilon^- &= \epsilon_{AB} \, \eta^A \otimes \epsilon^B, &\quad \text{in } D = 7.\\
\end{aligned}
\end{equation}
Using the definitions of~\cite{CSW3} for the fermionic fields, we find the internal components of the Killing spinor equations for the eleven-dimensional supersymmetry parameter $\varepsilon^-$ can then be neatly written in all dimensions as
\begin{equation}
\label{eq:AdS-susy-ferm}
\begin{aligned}
   \Big[ 
      \LC_m + \tfrac{1}{288} F_{n_1 \dots n_4} \left(
         \Gamma_m{}^{n_1 \dots n_4} 
         - 8 \delta_{m}{}^{n_1} \Gamma^{n_2 n_3 n_4}\right) 
       - \tfrac{1}{12} \tfrac{1}{6!} \tilde{F}_{mn_1 \dots n_6} 
	 \Gamma^{n_1 \dots n_6} 
      \Big] \varepsilon^- & = 0,\\
        \Big[ 
      \slashed{\LC} - \tfrac{1}{4} \slashed{F} 
      - \tfrac{1}{4} \slashed{\tilde{F}} 
      + \tfrac{D-2}{2} (\slashed{\der} \Delta) 
      \Big] \varepsilon^- 
      + \tfrac{D-2}{2} \Lambda \varepsilon^+ & = 0,
\end{aligned}
\end{equation}
where  $F, \,\tF$ are the internal four- and seven-form fluxes respectively, $\Gamma_{m}$ are now  $\Cliff(10,1)$ gamma matrices, and we define $\varepsilon^+$ by
\begin{equation} \label{eq:epsilon-plus}
\begin{aligned}
	\varepsilon^+ &= \ee^{-2\ii\theta} \eta_+ \otimes \epsilon^* 
		+ \ee^{2\ii\theta} \eta_- \otimes \epsilon, &\quad \text{in } D = 4,\\
	\varepsilon^+ &= -M_{AB} \, \eta^A 
		\otimes ( \ii \gamma^{(6)} \epsilon^B), &\quad \text{in } D = 5,\\
	\varepsilon^+ &= N_{AB} \, \eta_+^A \otimes \epsilon_2^B 
		+ (N^{-1})_{AB} \, \eta_-^A \otimes \epsilon_1^B, &\quad \text{in } D = 6,\\
	\varepsilon^+ &= \epsilon_{AB} \, \eta^A \otimes (\gamma^{(4)}\epsilon^B), 
		&\quad \text{in } D = 7.\\
\end{aligned}
\end{equation}
with $M_{AB} = \epsilon_{AC} M^C{}_B$. The reason for the choice of the superscripts $\varepsilon^{\pm}$ is that, as we will discuss momentarily (see also~\cite{CSW3}), they can be viewed as conjugate representations of $\Spin(D-1,1)\times\dHd$. Similar variables were identified in the earlier works~\cite{Martelli:2003ki,Gauntlett:2004zh}. 

In the following we can actually skip the discussion of $D = 6, \,d=5$ since we are only interested in backgrounds with minimal supersymmetry, and there is no such compactification to AdS${}_6$~\cite{Nahm:1977tg}. 
 Note that this is perfectly compatible with our generalised intrinsic torsion analysis -- we can see in table~\ref{table} that $D=6$ is the only case where the torsion contains no singlets, and thus cannot possibly accommodate the cosmological constant. We will discuss backgrounds with higher supersymmetry in a forthcoming paper~\cite{future}.

We can now decompose the full eleven-dimensional Killing spinor equation to obtain the conditions on the internal spinor, and thus on the geometry of the internal manifold. It is convenient at this point to make a choice of R-symmetry frame for the external spinors. This allows us write the internal conditions in terms of complex internal spinors, but breaks the external R-symmetry. For the $D=4$ case, we take the R-symmetry frame with $\theta = 0$ and simply write the equations for the complex internal spinor $\epsilon$ and do not write the conjugate equations for $\bar\epsilon$. For the $D=5$ case, we perform an $\SU(2)$ rotation to diagonalise the matrix $M^A{}_B$ to become $\sigma^3$. We may then write equations for the first ``half" $\epsilon \equiv \epsilon^1$ of the symplectic Majorana spinor $\epsilon^A$, but omit those for $\epsilon^2$, which follow by conjugating those for $\epsilon^1$. Similarly, we choose to write equations also only for $\epsilon \equiv \epsilon^1$ in the $D=7$ case, though this time we do not have to make any choice of R-symmetry frame to do so.

Decomposing~\eqref{eq:AdS-susy-ferm} thus leads to the internal equations
\begin{equation}\label{eq:ads-kse-int}
\begin{aligned}
\LC_m \epsilon + \tfrac{1}{288} 
   (\gamma_m{}^{n_1 \dots n_4} - 8 \delta_{m}{}^{n_1} \gamma^{n_2 n_3 n_4}) 
   F_{n_1 \dots n_4} \epsilon  - \tfrac{1}{12} \tfrac{1}{6!} \tilde{F}_{mn_1 \dots n_6} 			\gamma^{n_1 \dots n_6} = 0, \quad & \text{in } d = 4,6,7,\\
\slashed{\LC} \epsilon + (\slashed{\der} \Delta) \epsilon - \tfrac{1}{4} \slashed{F}\epsilon - \tfrac{1}{4} \slashed{\tilde{F}} \epsilon  +\Lambda\epsilon^* = 0,\quad & \text{in } d = 7,\\
\slashed{\LC} \epsilon + \tfrac{3}{2}(\slashed{\der} \Delta) \epsilon - \tfrac{1}{4} \slashed{F}\epsilon - \tfrac{1}{4} \slashed{\tilde{F}} \epsilon  -\tfrac32\Lambda\ii\gamma^{(6)}\epsilon = 0,\quad & \text{in } d = 6,\\
\slashed{\LC} \epsilon + \tfrac{5}{2}(\slashed{\der} \Delta) \epsilon - \tfrac{1}{4} \slashed{F}\epsilon - \tfrac{1}{4} \slashed{\tilde{F}} \epsilon  +\tfrac52\Lambda\gamma^{(4)}\epsilon = 0,\quad & \text{in } d = 4.\\
\end{aligned}
\end{equation}
These are then the AdS background Killing spinor equations we wish to examine.

As for the Minkowski case, these equations imply (see e.g.~\cite{Lukas:2004ip,Gauntlett:2004zh} for the cases of $d=6,7$) that the internal spinor is normalised as
\begin{equation}
\label{eq:spinor-norm}
	|| \epsilon ||{}^2 = \epsilon^\dagger \epsilon = \ee^\Delta.
\end{equation}

\subsubsection{Generic form in generalised geometry}

Let us start by rewriting these in the compact language of generalised geometry, which makes their larger local $\dHd$ symmetry manifest. For notational convenience, a first step is to introduce the rescaled spinor variables which are more naturally adapted to the $\dHd$ symmetry~\cite{CSW3}
\begin{equation}
\begin{aligned}
	\hat{\varepsilon}^\pm = \ee^{-\Delta/2} {\varepsilon}^\pm ,
\end{aligned}
\end{equation}
for the eleven-dimensional spinors and
\begin{equation}
\begin{aligned}
	\hat\epsilon^- = \ee^{-\Delta/2} \epsilon ,
\end{aligned}
\end{equation}
for the internal spinors.

A subtle point to note here is that, as is discussed in appendix \textbf{B} of~\cite{CSW3}, there are generically two ways of realising $\dHd$ in $\Cliff(d;\bbR)$, related by taking $\gamma^a\rightarrow -\gamma^a$, leading to two generically inequivalent spinor bundles $S^+$ and $S^-$. For instance in $d=7$, we have $\tilde{H}_7=SU(8)$ and $S^-$ is associated to the $\rep{8}$ of $SU(8)$ and $S^+$ to the $\rep{\bar{8}}$. In even dimensions these are actually isomorphic, $S^+\simeq S^-$, with the isomorphism given by the top gamma $\gamma^{(d)}$, for example in $d=4$ we have that $\hat\epsilon^+=\gamma^{(4)}\hat\epsilon^-$. With this in mind, we can then introduce torsion-free generalised spin-connections $D$ and find that the Killing spinor equations for AdS backgrounds~\eqref{eq:ads-kse-int} become, in manifestly $\dHd$-invariant form and for all dimensions
\begin{equation}\label{eq:gen-kse}
\begin{aligned}
	\Dgen \proj{J^-} \hat\epsilon^- &= 0 ,\\
	\Dgen \proj{S^+} \hat\epsilon^- &= -\tfrac{9-d}{2} \Lambda \hat\epsilon^+,
\end{aligned}
\end{equation}
where $\proj{X}$ denotes projection to the $X$ representation and $J$ is the representation of the vector-spinor in $d$-dimensions. 
We list the precise forms of these generic equations in the next section.

For completeness, we note that one can also write~\eqref{eq:gen-kse} in terms of undecomposed eleven-dimensional spinors. The group $\Spin(D-1,1)\times\dHd$ can be embedded in $\Cliff(10,1;\bbR)$, again in two different ways related by the overall sign of the gamma matrices. Labelling the representations corresponding to the spinors $\hat{S}^\pm$ and those for the vector-spinors $\hat{J}^\pm$, as in~\cite{CSW3}, we have
\begin{equation}\label{eq:gen-kse-11d}
\begin{aligned}
	\Dgen \proj{\hat{J}^-} \hat\varepsilon^- &= 0 ,\\
	\Dgen \proj{\hat{S}^+} \hat\varepsilon^- &= -\tfrac{9-d}{2} \Lambda \hat\varepsilon^+.
\end{aligned}
\end{equation}

\subsection{Generalised structures with singlet torsion}

The result now follows almost immediately. In~\cite{CSW4}, it was shown that the left-hand side of~\eqref{eq:gen-kse} matches exactly the intrinsic torsion $T_{\text{int}}$ of the generalised $G$-structure\footnote{Strictly, this holds assuming that the Killing spinor $\hat\epsilon^-$ has constant norm. However, this is always true by~\eqref{eq:spinor-norm}.} as listed in table~\ref{table}. The right-hand side, which simply vanished in the Minkowski case, now contains just the cosmological constant multiplying the ($G$-invariant) Killing spinor. Therefore, the Killing spinor equations are precisely equivalent to setting a singlet component of the intrinsic torsion to be proportional to $\Lambda$ and all other components to zero. In other words, the generalised connection that is compatible with the $G$-structure is not a torsion-free $D$ like in the Minkowski case, but $D+\Lambda$ instead, a connection with singlet torsion. One should thus think of the internal manifold as the generalised analogue of a manifold with weak special holonomy.

We can say a bit more about which singlet in the torsion in particular corresponds to the cosmological constant by looking in detail at each dimension. We skip $D=6$ since, as mentioned, there is no minimally supersymmetric AdS background there. 

The problem is simplified by noting that in~\eqref{eq:gen-kse} we find that the cosmological constant must lie in an $\dHd$ representation of the torsion that appears in the $S^+$ equation but not in the $J^-$ one (otherwise it would appear in the right-hand side of both equations).\footnote{For the $\dHd$ decomposition of the torsion representations see~\cite{CSW2}.} 

Another observation is that one expects the relevant singlet in the intrinsic torsion to break the R-symmetry from the $D$-dimensional Minkowski group to the $D$-dimensional AdS group. For Minkowski compactifications, the R-symmetry group arises as the commutant of the $G$ structure group inside $\dHd$. We therefore look to identify a singlet which transforms under this commutant group, exactly as in~\eqref{eq:ads-kse}, which is stabilised by the relevant AdS subgroup in the same way.

For $D=4$, the spinors $\hat\epsilon \equiv \hat\epsilon^-$ transform in the $\rep{8}$ of $\SU(8)$, while $\hat\epsilon^+ = \bar{\hat\epsilon}$ transform in the $\rep{\bar8}$. The Killing spinor equations become explicitly~\cite{CSW3,CSW4} 
\begin{equation}
\begin{aligned}
	(\Dgen \proj{J^-} \hat\epsilon^-){}^{[\alpha\beta\gamma]} &= \Dgen^{[\alpha\beta}\hat\epsilon^{\gamma]} = 0,\\
	(\Dgen \proj{{S}^+} \hat\epsilon^-){}_\alpha 
		&= -\Dgen_{\alpha\beta} \hat\epsilon^\beta = - \Lambda \bar{\hat\epsilon}_\alpha .
\end{aligned}
\end{equation}
The representations of the torsion which appear in the second equation but not the first are the $\rep{\bar{28}}+\rep{\bar{36}}$ (ie. objects with, respectively, two lower anti-symmetric and symmetric indices). However, the $\rep{\bar{28}}$ also appears in the conjugate gravitino variation $\Dgen_{[\alpha\beta}\bar{\hat\epsilon}_{\gamma]} = 0$, so this cannot contain the cosmological constant term. The Killing spinor is stabilised by an $\SU(7)\subset\SU(8)$ subgroup and the decomposition of the $\rep{\bar{36}}$ contains a singlet, so this must be the cosmological constant. Now we note that the commutant of  $\SU(7)$ in $\SU(8)$ is $U(1)$, and the singlet resulting from the symmetric two-index $\rep{\bar{36}}$ will carry a charge $\rep{2}$ under this $U(1)$. In fact, looking back at equations~\eqref{eq:AdS-susy-ferm} and~\eqref{eq:epsilon-plus}, one can see that this singlet is essentially $\ee^{-2 \ii \theta} \Lambda$ from equation~\eqref{eq:ads-kse}, but now viewed as transforming under the $U(1)$ commutant of $\SU(7)$. It will therefore be stabilised by a $\mathbb{Z}_2\subset U(1)$ subgroup, ie. precisely the R-symmetry group of $\mathcal{N}=1$ AdS${}_4$.

For $D = 5$, the generalised spin group is $\tilde{H}_6=\Symp(8)$, with spinors transforming in the $\rep{8}$. In $\Symp(8)$ indices,  the Killing spinor equations are explicitly
\begin{equation}\label{eq:gen6-kse}
\begin{aligned}
	(\Dgen \proj{J^-} \hat\epsilon^-){}^{[\alpha\beta\gamma]}  
		&= \Dgen^{[\alpha\beta}\hat\epsilon^{\gamma]} 
		+ \tfrac13 C^{[\alpha\beta} \Dgen^{\gamma]}{}_\delta \hat\epsilon^\delta = 0,\\
	(\Dgen \proj{{S}^+} \hat\epsilon^-){}_\alpha 
		&= \Dgen_{\alpha\beta} \hat\epsilon{}^\beta 
		= \tfrac{3}{2} \Lambda \hat\epsilon_\alpha ,
\end{aligned}
\end{equation}
where $C^{\alpha\beta}$ is the symplectic invariant. The only $\Symp(8)$-irreducible component of torsion that constrains the second equation but not the first transforms in the $\rep{36}$.

The special holonomy group here is $\Symp(6)\subset \Symp(8)$, and the decomposition of the $\rep{36}$ does indeed contain three singlets, one of which which we can thus identify as the cosmological constant. The commutant subgroup in turn is $\Symp(2)\simeq SU(2)$, and the $\Symp(6)$ singlets transform in the adjoint $\rep{3}$ of $\SU(2)$. Note however, that the $\Symp(6)$ structure stabilises not only one spinor $\hat\epsilon$, but also a second spinor $\hat\epsilon'$. These two spinors form a symplectic Majorana pair $\hat\epsilon^A$. However, here we can treat this index as the $\Symp(2)$ index labelling the two spinors preserved by $\Symp(6)$. In a generic $\Symp(2)$ frame, we can then rewrite the second line of~\eqref{eq:gen6-kse} as
\begin{equation}
		\Dgen_{\alpha\beta} (\hat\epsilon^A){}^\beta 
		= \tfrac{3}{2} \Lambda M^A{}_B (\hat\epsilon^B){}_\alpha ,
\end{equation}
with the constant matrix $M^A{}_B$ as in~\eqref{eq:ads-kse}. As there, the cosmological constant comes attached with the traceless Hermitian matrix $M^A{}_B$, and so is naturally an element of the triplet representation. Fixing this element will therefore break the $\SU(2)$ down to $U(1)$, the R-symmetry group of $\mathcal{N}=1$ AdS${}_5$.

Finally, in $D = 7$ spinors transform in the $\rep{4}$ of $\Spin(5)\simeq \Symp(4)$. The Killing spinor equations can be written explicitly~\cite{CSW3} as
\begin{equation}
\begin{aligned}
(\Dgen \proj{{J}^-} \hat\epsilon^-) &= 2(\gamma^j \Dgen_{ij} \hat\epsilon 
         - \tfrac15 \gamma_i \gamma^{jj'} \Dgen_{jj'} \hat\epsilon ) = 0,\\
	(\Dgen \proj{{S}^+} \hat\epsilon^-) &= -\gamma^{ij}\Dgen_{ij} \hat\epsilon 
		= -\tfrac{5}{2} \Lambda \hat\epsilon ,
\end{aligned}
\end{equation}
where we are actually using the more familiar $SO(5)$ indices $i,j\dots$ and omitting spinor indices. The only component of the torsion that appears just in the second equation is a singlet of $\Spin(5)$. This obviously is still invariant under the Killing spinor stabiliser $\Symp(2)\subset\Symp(4)$ so it must be the cosmological constant. Clearly the singlet is also automatically invariant under the entire commutant subgroup $\Symp(2)$, which indeed is the R-symmetry group of AdS${}_7$.  

We should note that in M theory there are actually no smooth AdS${}_7$ backgrounds which are strictly $\mathcal{N}=1$~\cite{Acharya:1998db}. However, recently a family of genuinely $\mathcal{N}=1$ solutions in massive Type IIA theory was discovered~\cite{Apruzzi:2013yva}. To describe these backgrounds as generalised structures, one would presumably need a slightly modified formulation of $\Edd\times\bbR^+$ generalised geometry for the massive Type IIA theory, which would be beyond the scope of this paper. For M theory the only possibility is the $S^4$ solution with maximal $\mathcal{N}=2$ supersymmetry. We will discuss backgrounds with higher preserved supersymmetry in a subsequent paper~\cite{future}, but we remark that in this case the generalised structure group reduces to the identity, with the commutant being the entire $\Symp(4)$. 
The singlet in the torsion that we identified above would not break this commutant group (since it is a singlet of $\tilde{H}_4=\Symp(4)$), reflecting that the AdS R-symmetry is the full $\USp(4)$ group for $\mathcal{N}=2$.
The generalised parallelisation on AdS${}_7\times S^4$ is presented in detail in~\cite{Lee:2014mla}, as an example of the generic appearance of this structure in maximally supersymmetric compactifications.


\section{Discussion}

We have shown that spaces admitting the appropriate generalised $G$-structure with constant singlet torsion are precisely the minimal AdS flux backgrounds. This is summarised in table~\ref{tab:AdS}.
\begin{table}[htb]
\centering
\begin{tabular}{lllll}
	$d$ & $G$ & $G_{\text{com}}$ & R-symmetry &  $\TTint$  \\ 
	\hline 
	7 & $\SU(7)$ & $ U(1)$ &  $\bbZ_2$ & $\rep{1}_2$ \\
	6 & $ \Symp(6)$ &$ \USp(2)$ &  $U(1)$ & $\repp{3}{1}$ \\
	5 & $\Symp(2)\times\Symp(4)$ & $\USp(2)$ & ---
             & no singlets \\
	4 & $\Symp(2)$ & $\USp(2)$  &$\USp(2)$ & $\repp{1}{1}$
\end{tabular}
\caption{Generalised structure subgroups $G\subset \dHd$, commutant groups $G_{\text{com}}$ of $G$ in $\dHd$, AdS R-symmetry groups and non-vanishing generalised intrinsic torsion as representations of $G_{\text{com}}\times G$ for minimal supersymmetry in AdS backgrounds.}
\label{tab:AdS}
\end{table}

We stress again that even though we have focused on the M theory case, the formalism is such that the results necessarily extend to Type II strings.

As a corollary, we remark that there exists a definition of generalised Ricci curvature~\cite{CSW2}, which given an arbitrary spinor $\epsilon$ reads schematically (see~\cite{CSW3} for explicit definitions)
\begin{equation}
\begin{aligned}
\GenRic\cdot\epsilon = \Dgen \proj{J} (\Dgen \proj{J}\epsilon)+\Dgen \proj{J} (\Dgen \proj{S}\epsilon),\\
\GenS \epsilon = \Dgen \proj{S} (\Dgen \proj{J}\epsilon)+\Dgen \proj{S} (\Dgen \proj{S}\epsilon),
\end{aligned}
\end{equation}
where $\GenS$ is the generalised Ricci scalar and $\GenRic$ is the traceless part of the generalised Ricci. The vanishing of the full generalised Ricci corresponds to the Minkowski equations of motion~\cite{CSW2}. Using the result of appendix \textbf{B} of~\cite{CSW4}, it is then easy to show that if the intrinsic torsion conditions~\eqref{eq:gen-kse} hold, we have that
\begin{equation}
\GenRic = 0, \quad \GenS = (9-d)(10-d)\Lambda^2 ,
\end{equation}
ie. we have the natural generalised analogue of the Einstein manifold condition. This is simply a more geometric (and much simpler to derive) restatement of the supergravity result that in a supersymmetric AdS background the equations of motion are satisfied~\cite{Gauntlett:2002fz}. On the other hand, if we had not assumed that the singlet torsion $\Lambda$ was a constant but rather an arbitrary function, these equations would not hold. However, imposing them would then force $\Lambda$ to be constant~\cite{Lukas:2004ip}. 

Another result that naturally generalises in this setting is of that of the cone spaces over the classical Sasaki--Einstein and weak $G_2$ spaces -- well-known AdS backgrounds -- which then become special holonomy manifolds. The same will happen here, now for generic backgrounds. Viewing the $D$-dimensional AdS space as a warped product of $(D-1)$-dimensional flat space and a line, implies that the cones over the spaces listed in table~\ref{tab:AdS} must all be special holonomy spaces for $E_{d+1(d+1)}\times\bbR^+$ generalised geometry~\cite{CSW4}.  

An interesting avenue of further study would be to see if the same statements can be made in generalised geometries other than ones based on the exceptional $\Edd$ groups, or whether any such statements can be made for $D\leq3$. Recent work~\cite{AdS3foliations} showed that generic $\mathcal{N} =1$ AdS$_3$ solutions of M theory can be described as foliations of seven-dimensional spaces, and it would be interesting to make contact with that picture using generalised geometry.
For example, one could examine the conditions for backgrounds with zero internal $F_{(4)}$ flux in $\Spin(8,8)\times\bbR^+$ generalised geometry~\cite{CSC}.

The most obvious extension of this work, however, would be to demonstrate an analogous statement for all supersymmetric AdS backgrounds of M and Type II theories, not just the ones with minimal supersymmetry. In forthcoming work~\cite{future} we will show how the methods used in~\cite{CSW4} for the Minkowski case can be expanded to deal with higher $\mathcal{N}$ backgrounds, at which point one could hope the classification of AdS will follow similarly to that outlined in this paper.


\begin{acknowledgments}
We would like to thank Dan Waldram for helpful discussions.
C.~S-C.~is supported by a grant from the Foundational Questions Institute (FQXi) Fund, a donor advised fund of the Silicon Valley Community Foundation on the basis of proposal FQXi-RFP3-1321 (this grant was administered by Theiss Research). 
A.~C.~is supported by the German Research Foundation (DFG) within the Cluster of Excellence ``QUEST''. C.~S-C.~would like to thank Imperial College London for hospitality and also the EPSRC Programme Grant EP/K034456/1 ``New Geometric Structures from String Theory" for visitor support.
\end{acknowledgments}



\end{document}